\documentclass[useAMS,usenatbib]{mn2e}

\usepackage{graphicx,subfigure,color}

\title[Automatic data analysis for SBM]{Automatic data analysis for Sky Brightness Monitor}
\author[Zhao et al.]{M. Y.
Zhao$^{1}$\thanks{E-mail:myzhao@ynao.ac.cn}, Y.
Liu$^{1,2}$\thanks{E-mail:lyu@ynao.ac.cn}, A. Elmhamdi$^{1,2}$, A.
S. Kordi$^{2}$, H.A. Al-trabulsy$^{2}$,
\newauthor
X. F. Zhang$^{1}$, T. F. Song$^{1}$, S. Q. Liu$^{1}$, Y. D.
Shen$^{1}$, Z. J. Tian$^{1}$ and Y. H. Miao$^{1}$\\
$^{1}$Yunnan Observatories, Chinese Academy of Sciences, Kunming, 650011, China\\
$^{2}$Department of Physics and Astronomy, King Saud University,
P.O. Box 2455, 11451, Saudi Arabia}
\begin{document}

\date{Accepted . Received ; in original form }

\pagerange{\pageref{firstpage}--\pageref{lastpage}} \pubyear{2014}

\maketitle

\label{firstpage}

\begin{abstract}
The Sky Brightness Monitor (SBM) is an important instrument to
measure the brightness level for the sky condition, which is a
critical parameter for judging a site for solar coronal
observations. In this paper we present an automatic method for the
processing of SBM data in large quantity, which can separate the
regions of the Sun and the nearby sky as well as recognize the
regions of the supporting arms in the field of view. These processes
are implemented on the data acquired by more than one SBM
instruments during our site survey project in western China. An
analysis applying the result from our processes has been done for
the assessment of the scattered-light levels by the instrument.
Those results are considerably significant for further
investigations and studies, notably to derive a series of the other
important atmospheric parameters such as extinctions, aerosol
content and precipitable water vapor content for candidate sites.
Our processes also provide a possible way for full-disk solar
telescopes to track the Sun without an extra guiding system.
\end{abstract}


\section[]{Introduction}
The sky brightness is a critical parameter for judging a potential
site for direct coronal observation. Due to the large difference in
the brightness of the solar disk and its nearby sky (halo), it is
difficult to achieve the accurate and direct measurement of the sky
brightness during the day time. The corona can only be observed
during total solar eclipses until the {\it Lyot} coronagraph was
invented \citep{Lyot1930, Lyot1939}. Afterwards, a photometer for
measurement of sky brightness near the Sun is constructed by
\citet{Evans1948}. Evans Sky Photometer (ESP) has been used for a
long time at various solar observatories \citep{Garcia1983,
Labonte2003}. A modern Sky Brightness Monitor (SBM) has been
developed for the Advanced Technology Solar Telescope (ATST) site
survey project, which can image the solar disk and the nearby sky
simultaneously at 450, 530, 890 and 940 nm \citep{Penn2004,
Lin2004}. In addition, the SBM data can yield information about the
extinction, the aerosols, and the precipitable water vapor content
of the atmosphere.

To meet the future development of the Chinese Giant Solar Telescope
(CGST) and large-aperture coronagraphs projects, our SBM is designed
and developed following the same idea from \cite{Lin2004}, as one
fundamental tool \citep{Liu2012b}. These SBMs have been applied at
many different candidate sites up to now \citep{Liu2011, Liu2012a,
Liu2012b}. The optical configuration of one SBM consists of an
external occulter, an objective lens, a neutral density filter on
the objective lens, and the four bandpass filters \citep{Lin2004}.
Generally, the original images acquired by the SBM are not
intuitional and concise for a directly further analysis. Moreover,
the absence of an auto guide system to serve the equatorial tracking
system would result in an indeterminateness of the positions of the
solar disk in the SBM's images. \citet{Wen2012} studied and compared
the possible methods for determining the solar disk center in SBM's
data. However, they could not break away from the constraint of some
experienced parameters which need personnel intervention.
Furthermore, the invalid regions for calculating the sky brightness
are excluded manually in former work, too\citep{Liu2012b}. With the
rapid progress of our site survey project, the accumulation of huge
amounts of data urgently necessitates an efficient, accurate and a
fully automated processing approach to deal with the SBM data.

For this purpose, we have developed a method for automatically
processing the SBM data, which allows separating the regions of the
Sun and the nearby sky as well as permits to automatically recognize
the regions of the supporting arms. In the next section, we describe
the main adopted technologies for our procedure. In Section 3, we
present the data reduction in detail. In Section 4, we apply our
procedure on a set of SBM's data and analyze the scattered light
based on the results of our procedure. Our conclusions are presented
and highlighted in Section 5.

\section[]{Main Technologies}
\begin{figure*}
\centering
\includegraphics[width=84mm,angle=90]{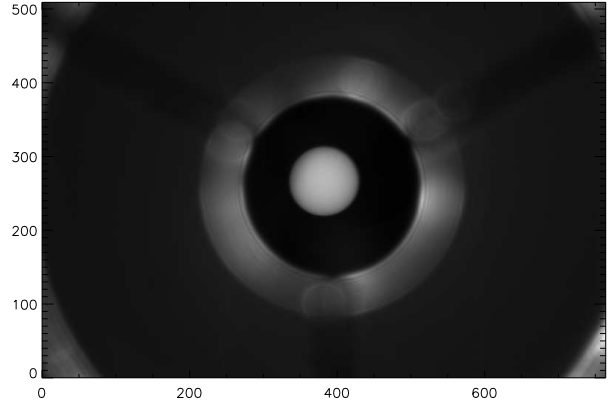}
\caption{A typical SBM image taken at {\it Gaomeigu} Station in
Lijiang city. X- and Y- directions represent the two dimensions of
the image in pixels, similarly hereafter.} \label{fig:1}
\end{figure*}

The SBM instruments can simultaneously image the Sun and its nearby
sky regions. The data is useful for directly deducing the normalized
sky brightness which is a critical parameter for coronagraph site
investigation. Fig.~\ref{fig:1} shows a typical SBM image taken at
{\it Gaomeigu} Station. Near the center of the image, the attenuated
solar disk and the projection of the ND4 (nominal optical density of
4) filter are shown. The bright ring is caused by the diffraction at
the edge of the ND4 filter, while the three narrow strips are from
the projections of the supporting arms of the ND4 filter.

The brightness of the sky within a few solar radii away from the
solar center at a good coronal site is expected to be below $10^{-5}
\times I_{\odot}$, where $I_{\odot}$ is the intensity of the solar
disk \citep{Lin2004}. It is convenient to measure the sky brightness
with millionths of $I_{\odot}$. In order to extract the regions of
the solar disk and the Sun's nearby sky, it is necessary to
accurately determine the solar radius and the position of the solar
center in the image. Because the region of the Sun's nearby sky on
the image contains the projections of the supporting arms which must
be excluded in the calculation, the accurate positions of these
projections are also needed for each image. In our procedure, there
are mainly three technologies used to achieve the separation of
these regions: image binarization with \textit{Otsu}'s method, image
matching with cross-correlation, and image fitting with the method
of least-squares.

In digital image processing, \textit{Otsu}'s method is usually used
to automatically perform clustering-based image binarization. The
algorithm assumes that the image contains two classes of pixels
corresponding to two peaks on its histogram (foreground and
background). The optimum threshold to separate those two classes is
calculated by minimizing their combined spread (intra-class
variance) \citep{Otsu1979}.

In \textit{Otsu}'s method, the intra-class variance (the variance within the class) is
defined as a weighted sum of variances of the two classes:
\begin{equation}
\sigma^2(T)=\omega_1(T)\sigma_1^2(T)+\omega_2(T)\sigma_2^2(T)
\label{eq:1}
\end{equation}
where $\omega_i$ are the probabilities of the two classes separated
by a threshold $T$ and $\sigma^2_ i$ are variances of these classes.

\begin{figure*}
\subfigure[]{ \centering
\includegraphics[width=84mm,angle=90]{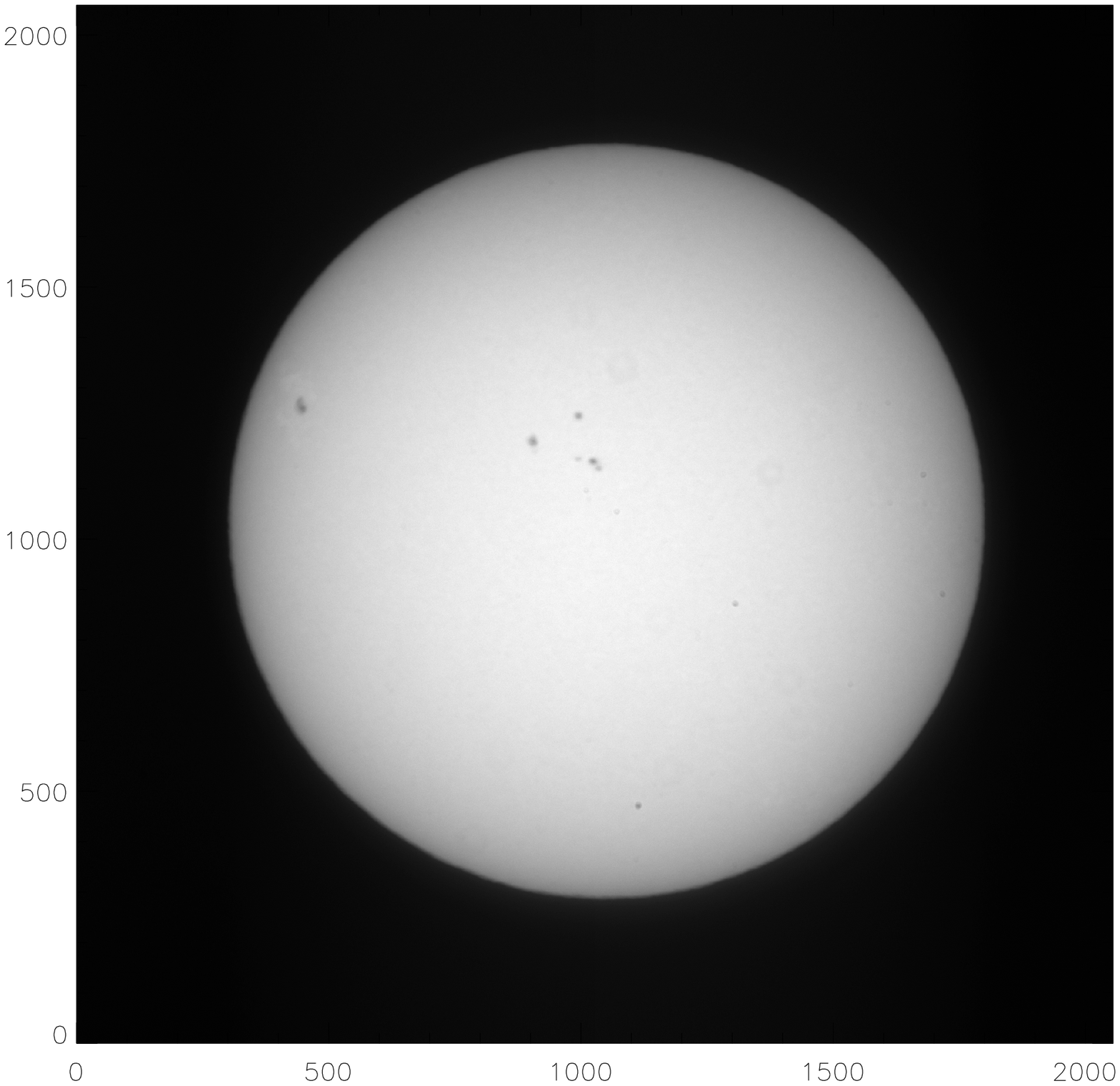}
}%
\subfigure[]{ \centering
\includegraphics[width=84mm,angle=90]{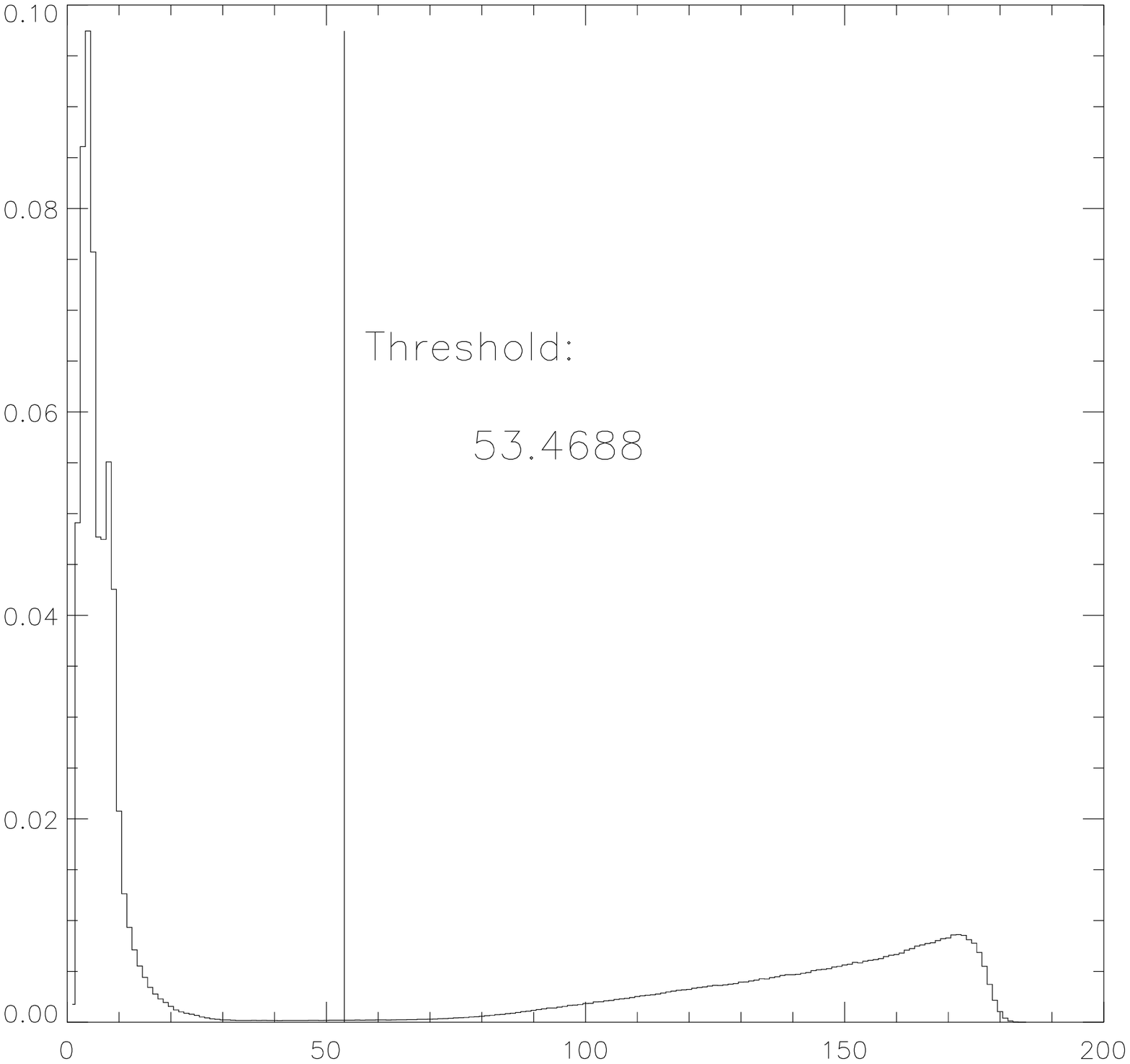}
}%
\\
\subfigure[]{ \centering
\includegraphics[width=84mm,angle=90]{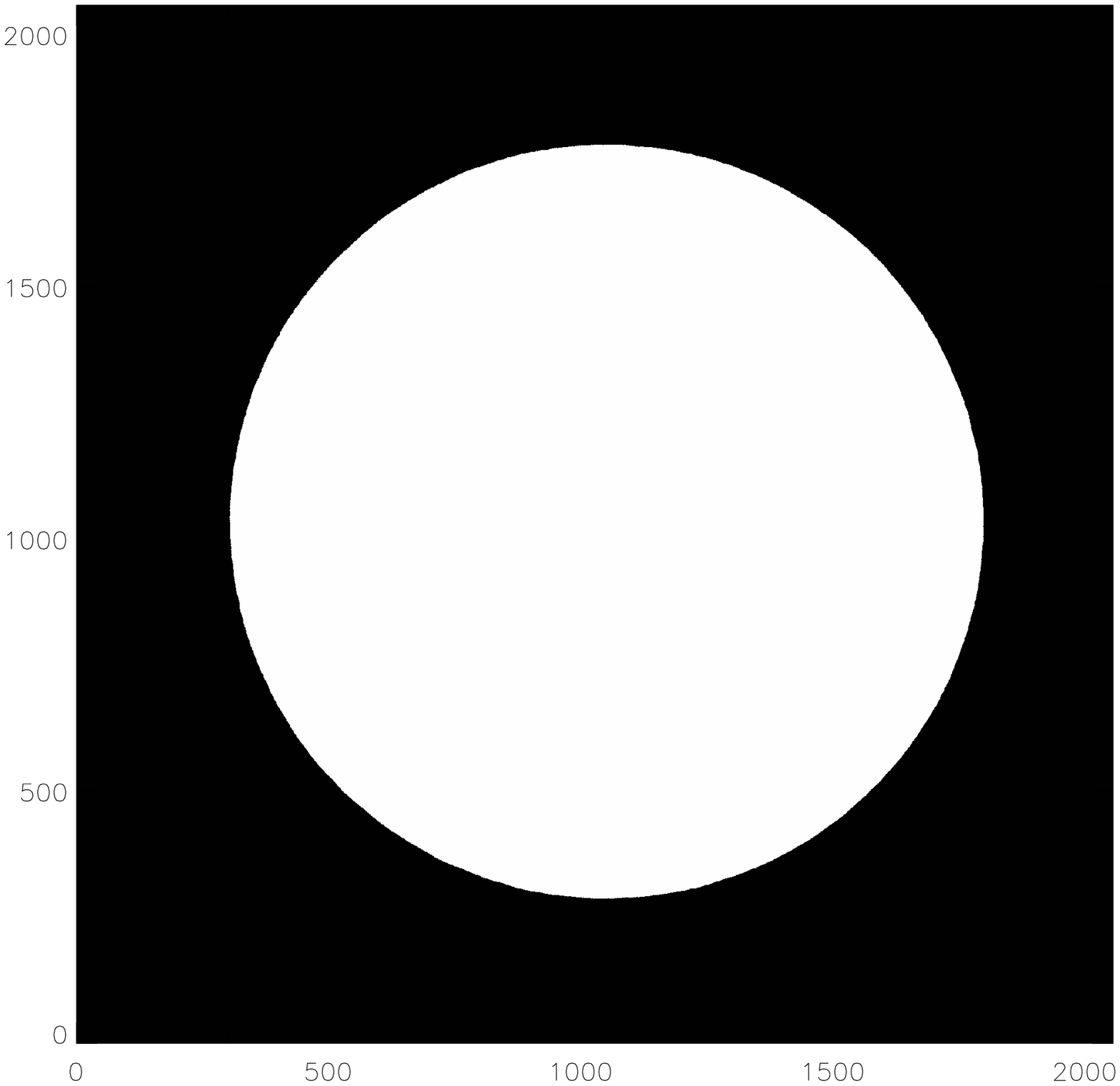}
}%
\subfigure[]{ \centering
\includegraphics[width=84mm,angle=90]{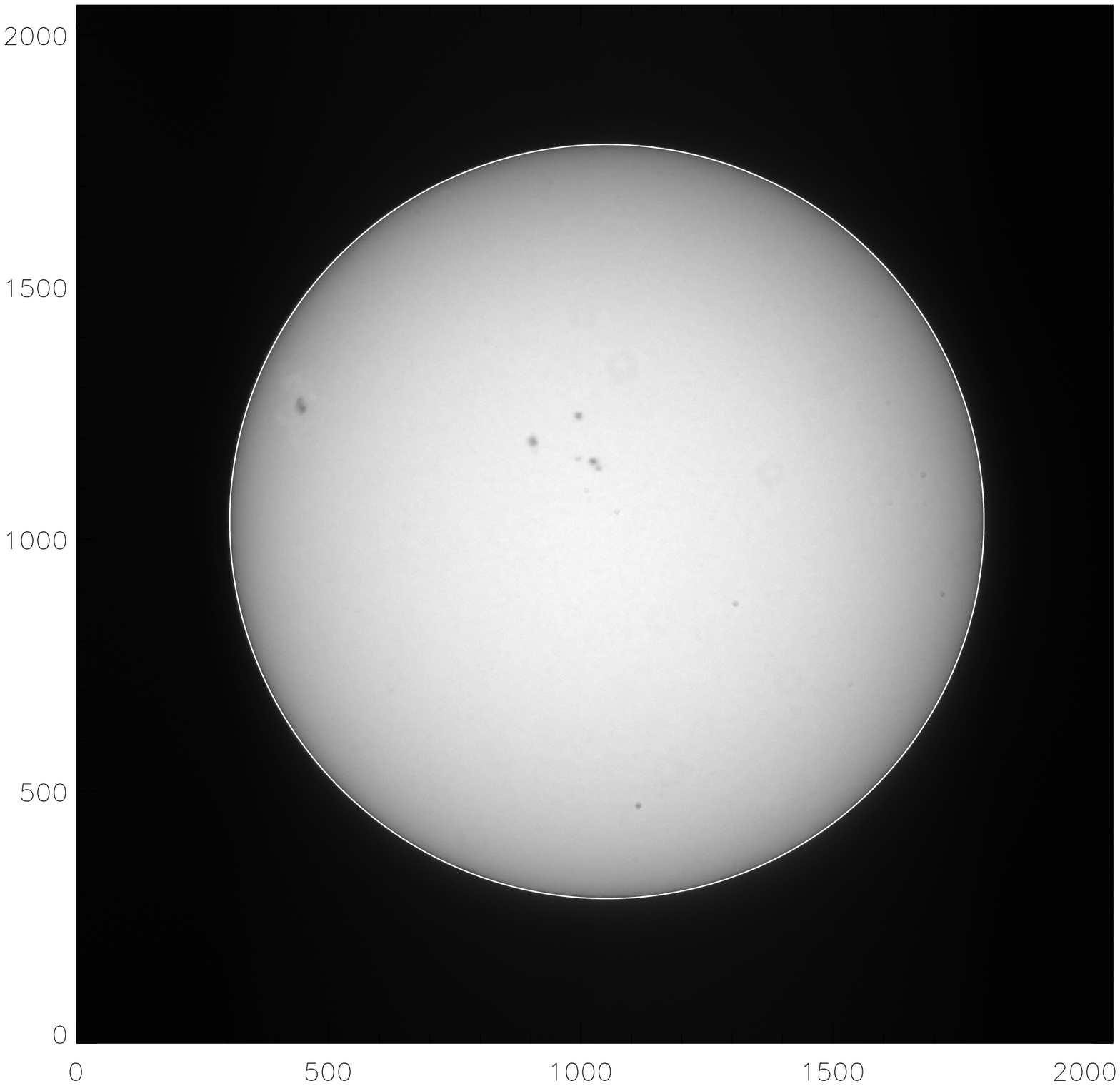}
}%
\caption{An illustration of image binarization based on {\it Otsu}'s
method. The image is taken at Full-Shine lake in Yunnan. (a) is a
whitelight solar image. (b) is the result applying Otsu's method on
(a). (c) is the binarized image based on the threshold from (b). (d)
shows the boundary of (c) superimposed on (a).} \label{fig:2}
\end{figure*}

In this step, the image is re-scaled to 8-bit and the normalized
histogram is calculated. Then we test all the possible thresholds T
from 0 to 255 to calculate the intra-class variances and find the
desired T in the 256 thresholds. At last, the actual threshold for
the origin image is obtained by an inverse scaling mapping. For
example, applying \textit{Otsu}'s method on a white light solar
image would locate the boundary with a high accuracy
(Fig.~\ref{fig:2}). In our procedure, \textit{Otsu}'s method is
mainly used for the edge detecting of the SBM image.

Correlation techniques is predominately used for template matching
in image processing \citep{Gonzalez2002}. If we want to determine
whether an image $f(x,y)$ containing a particular object or region
of interest, we let $h(x,y)$ be that object or region (template).
Then, if there is a match, the correlation of the two functions will
be maximum at the location where $h(x,y)$ finds a correspondence in
$f(x,y)$.

\begin{figure*}
\subfigure[]{ \centering
\includegraphics[width=84mm,angle=90]{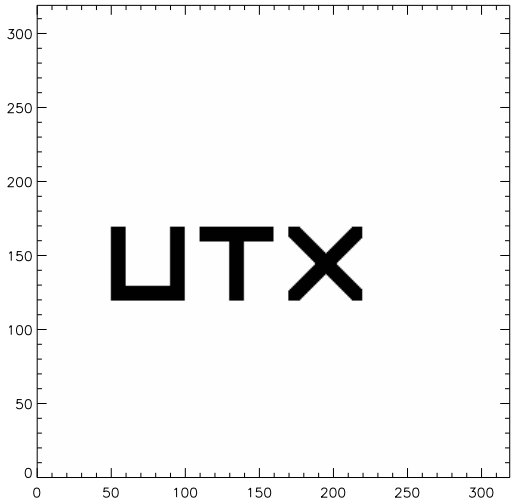}
}%
\subfigure[]{ \centering
\includegraphics[width=84mm,angle=90]{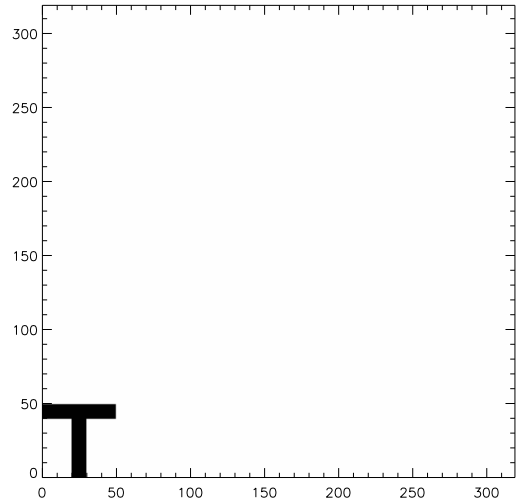}
}%
\\
\subfigure[]{ \centering
\includegraphics[width=84mm,angle=90]{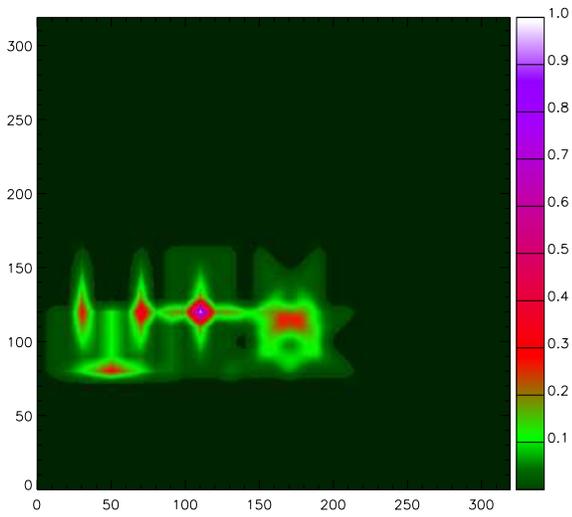}
}%
\subfigure[]{ \centering
\includegraphics[width=84mm,angle=90]{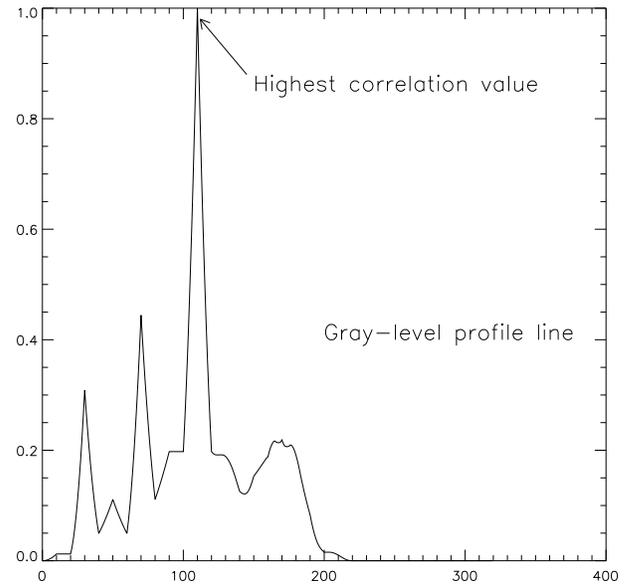}
}%
\caption{A simple illustration of image correlation. It is
constructed following the example 4.41 in \citet{Gonzalez2002}.(a)
is the image and (b) is the template. (c) shows the cross
correlation between (a) and (b). (d) is the horizontal profile line
through the highest value in (c).} \label{fig:3}
\end{figure*}

Cross-correlation of two continuous functions {\it f} and {\it h} is
defined as
\begin{equation}
f(t)\circ h(t)=\int_{-\infty}^{\infty}f^{\ast}(\tau)h(t+\tau)d\tau
\label{eq:2}
\end{equation}
where the asterisk denotes the complex conjugate of $f$. If $f$ and
$h$ are real valued functions, this formula essentially slides the
function $h$ a displacement to function $f$ and calculates the
integral of their product at each position. When the functions
match, the value of $f\circ h$ is maximized. This is because when
peaks (or troughs) are aligned, they make a large contribution to
the integral. With complex valued functions $f$ and $h$, taking the
conjugate of $f$ ensures that aligned peaks (or aligned troughs)
with imaginary components will contribute positively to the
integral. Therefore, one can use the cross-correlation to find how
much $h$ must be shifted to make it identical to $f$. An
illustrating example is shown in Fig.~\ref{fig:3}. In our procedure,
cross-correlation is used for edge extracting and to determine the
positions of the supporting arms.

The method least-squares (LSQ) is extensively utilized in data fitting. It is
a standard approach to the approximate solution of over-determined
systems, in which the number of measurements is more than that of
unknown variables. Least-squares means that it minimizes the sum of
the squared residuals. In our procedure, least-squares is used to
fit the circles by the points of the edges of the solar disk and the
ND4 filter.

\begin{figure*}
\centering

\includegraphics[width=84mm,angle=90]{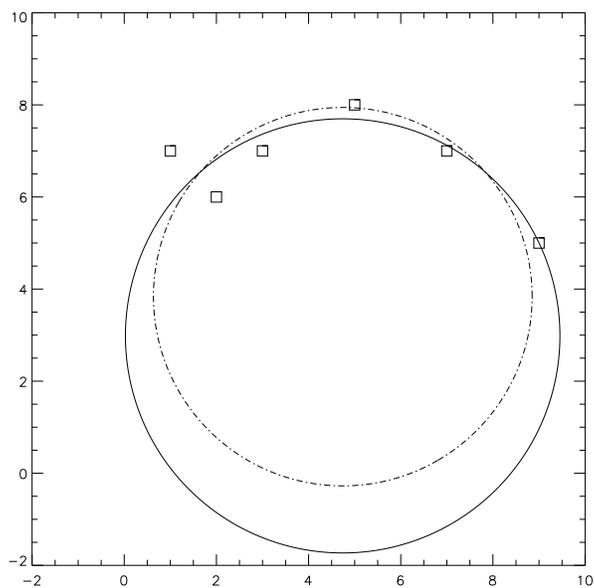}
 \caption{An example of
least-squares fitting. The dot-dashed line represents the algebraic
fitting and the solid line represents the geometric fit.}

\label{fig:4}
\end{figure*}

For a circle with the algebraic equation,
\begin{equation}
F(x_i,y_i,\vec{\beta})=a(x_i^2+y_i^2)+b_1x_i+b_2y_i+c=0,
\label{eq:3}
\end{equation}
determining the parameters ($\vec{\beta}=[a,b_1,b_2,c]$) of the
algebraic equation in the least-squares sense is called algebraic
fit, while minimizing the sum of the squares of the distances to the
center point
\begin{equation}
d_i^2=(\|\vec{z}-\vec{x}_i\|-r)^2
\label{eq:4}
\end{equation}
is called geometric fit, where $\vec{z}$ is the center
coordinate of the circle and $r$ is the radius of the circle
\citep{Gander1994}. The sum of the squared residuals of the distance
function is obviously a non-linear function. Therefore, we solve
this non-linear least-squares problem iteratively with the {\it
Gauss-Newton} method. And the algebraic solution is used as the
starting vector for the iteration. An example is shown in
Fig.~\ref{fig:4}. The results of the algebraic and the geometric fits are very different. For fitting analysis of data, the algebraic fit only tries to find the curve which minimizes the vertical displacement while the geometric fit tries to minimize the orthogonal distance. On this account, geometric fit sometimes is mentioned as "best fit", because it can provide more aesthetic and accurate results.

\section[]{Description of the Data Reduction}
The positions of the solar disk in a time sequence are crucial
parameters for the sky brightness calculations. Due to the actual
operation on the instrument, the solar image center is usually
shifted around the center of the ND4 filter. Because the two ND2
filters are fixed on the supporting arms, the position of the center
and the projection size of the filters are also important parameters
for the calculations.

\subsection{Data selection}
In actual observations, there may be a lot of data that are
mistakenly collected due to the observational constraints. Because
the SBM is a portable measurement designed for field observations,
the weather condition has a great influence on the acquired images.
Moreover, in the absence of a guiding system and high-precision
tracking system, the SBM often gathers bad images in which the Sun
moves out of the field of view. Hence a data selection program must
be performed prior to any further calculations.

\begin{figure*}
\subfigure[]{
\includegraphics[width=84mm,angle=90]{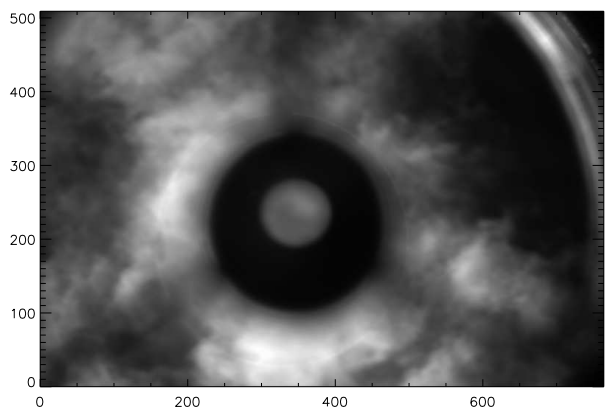}
}%
\subfigure[]{
\includegraphics[width=84mm,angle=90]{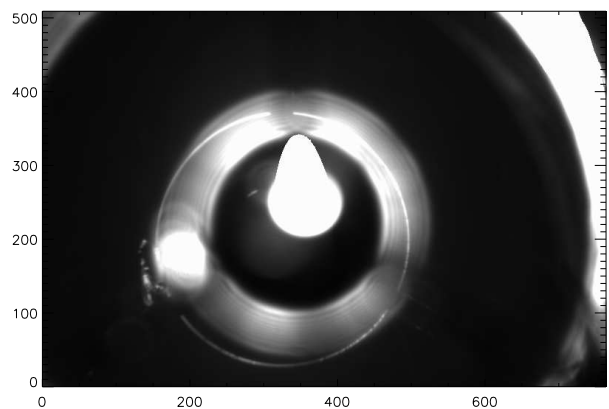}
}%
\caption{(a) represents the bad image covered by clouds. (b)
represents the bad image, clearly saturated.} \label{fig:5}
\end{figure*}

There are mainly two types of bad data in the observation which are
represented in Fig.~\ref{fig:5}. Fig.~\ref{fig:5} (a) represents the
bad image covered by the clouds and (b) shows the bad image which is
saturated. We use an operation on experiential thresholds to
recognize these situations. Binary operations are applied to the
images according to the thresholds by \textit{Otsu}'s method and
0.999 maximum, respectively. Then the areas of the binarization
images are calculated. Because \textit{Otsu}'s threshold separates
two typical classes of pixels optimally and the clouds attenuation
the differences of these two classes, the areas from \textit{Otsu}'s
threshold will be very large. And if the image is saturated, there
will be too many pixels with values identical to the maximum. Hence
the area from 0.999 maximum threshold will be too large. We note
that these two criteria are not independent. For example, when the
weather is seriously cloudy, the image will be saturated in large
area, and the binarization images based on \textit{Otsu}'s threshold
and 0.999 maximum threshold are nearly the same.

We also take into account some other ordinary situations such as for
the Sun totally out of the field of view. Our data selection
procedure is very effective to automatically get rid of those data.
This has been verified based on all the SBM's data, collected in
2013, during the solar site survey. However the accurate thresholds
can not be given because our criteria depend on the experience. In
fact, we adopt a slight strict threshold to ensure the selecting to
be sufficient while a very few regular data will be excluded
improperly.

\subsection{Edge detection and extraction}
\begin{figure*}

\subfigure[]{ \centering
\includegraphics[width=84mm,angle=90]{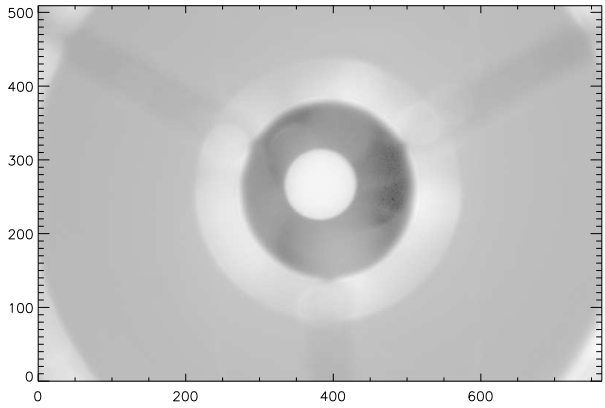}
}%
\subfigure[]{ \centering
\includegraphics[width=84mm,angle=90]{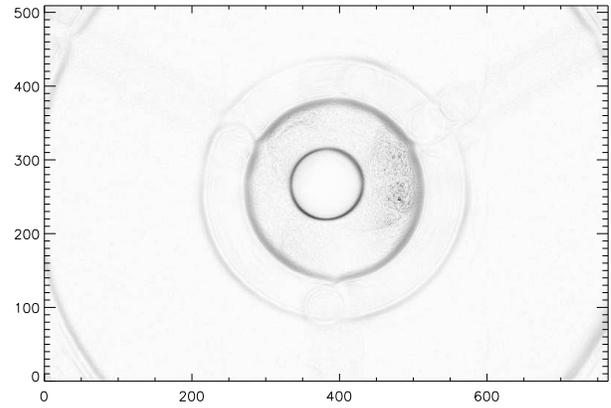}
}%
\\
\subfigure[]{ \centering
\includegraphics[width=84mm,angle=90]{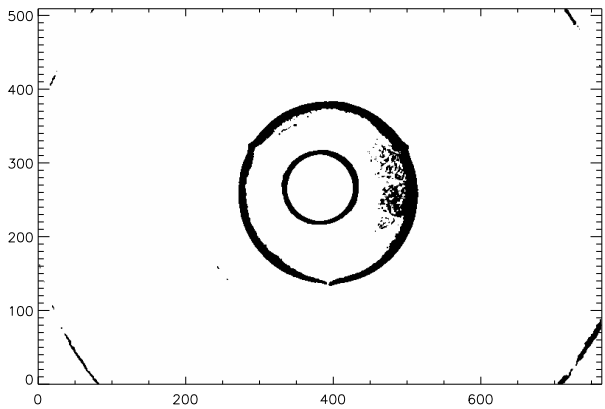}
}%
\subfigure[]{ \centering
\includegraphics[width=84mm,angle=90]{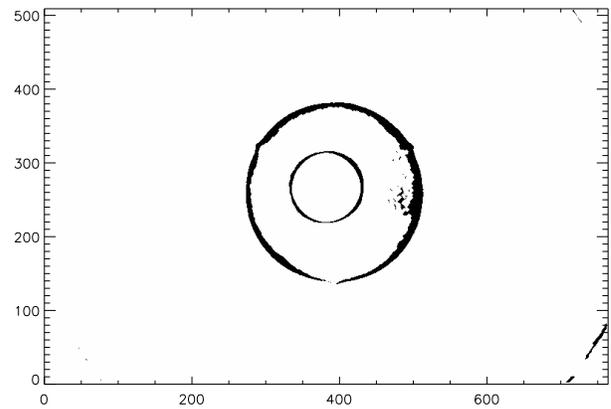}
}%
\caption{(a) is the enhanced image, which is the logarithm of the
initial image . (b) is the gradient of image (a) by \textit{Sobel}
operator. (c) is the binarized image based on the threshold from
(b). (d) is the eroded image of (c) by morphological method. We note
that the gray levels of the binarized images are inverted for
display, similarly hereafter.} \label{fig:6}
\end{figure*}

The second step in our procedure is edge detection. We apply {\it
Sobel operator} on the enhanced SBM image to get the gradient image
and use \textit{Otsu}'s Method to threshold the gradient image. Then
the edge of the solar disk combined with that of the ND4 filter and
some other small structures are obtained (see Fig.~\ref{fig:6}).
Here a morphological erosion with a circular structuring element of
$7\times 7$ pixels is employed to make the edge more distinct.

\begin{figure*}
\subfigure[]{ \centering
\includegraphics[width=84mm,angle=90]{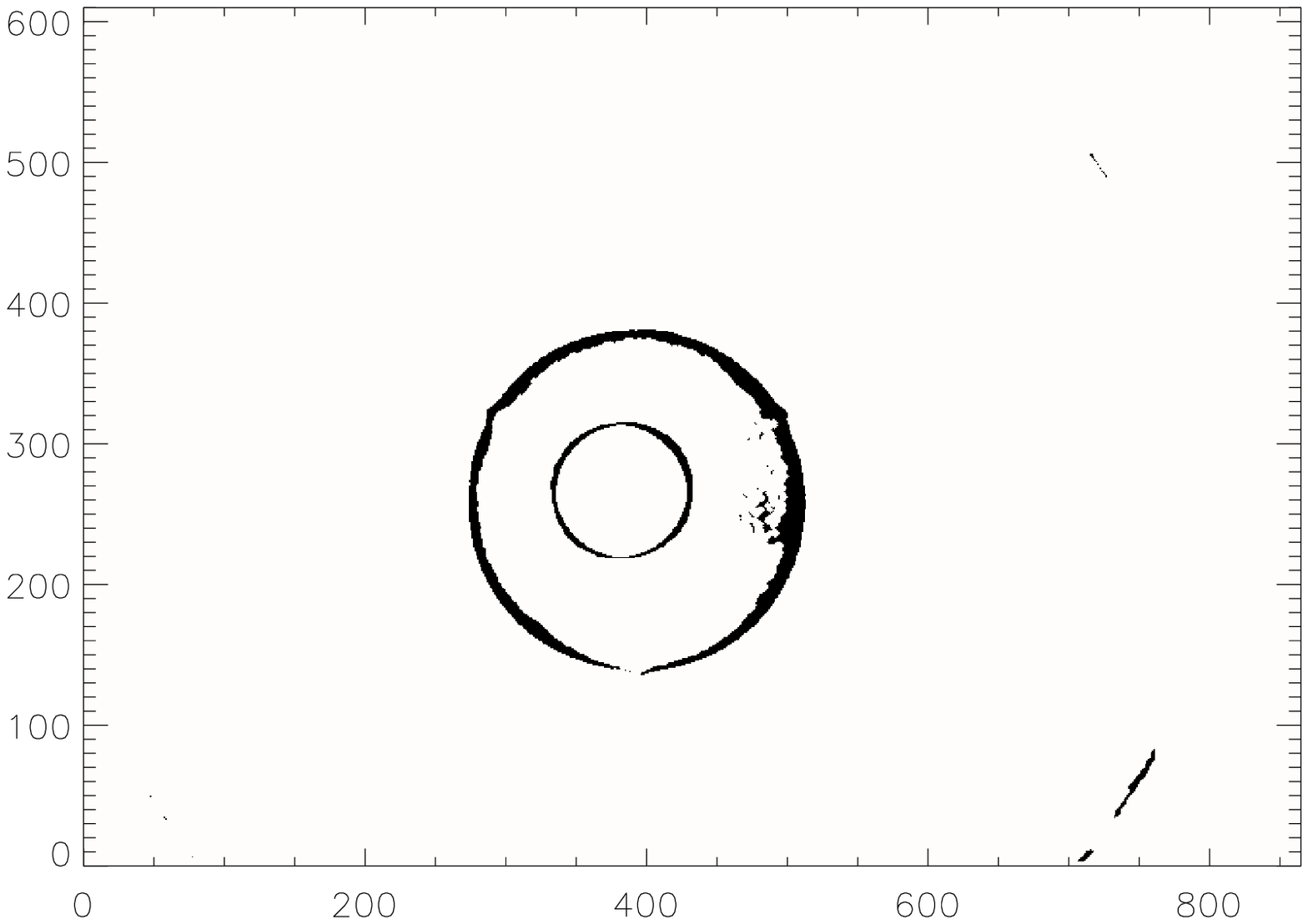}
}%
\subfigure[]{ \centering
\includegraphics[width=84mm,angle=90]{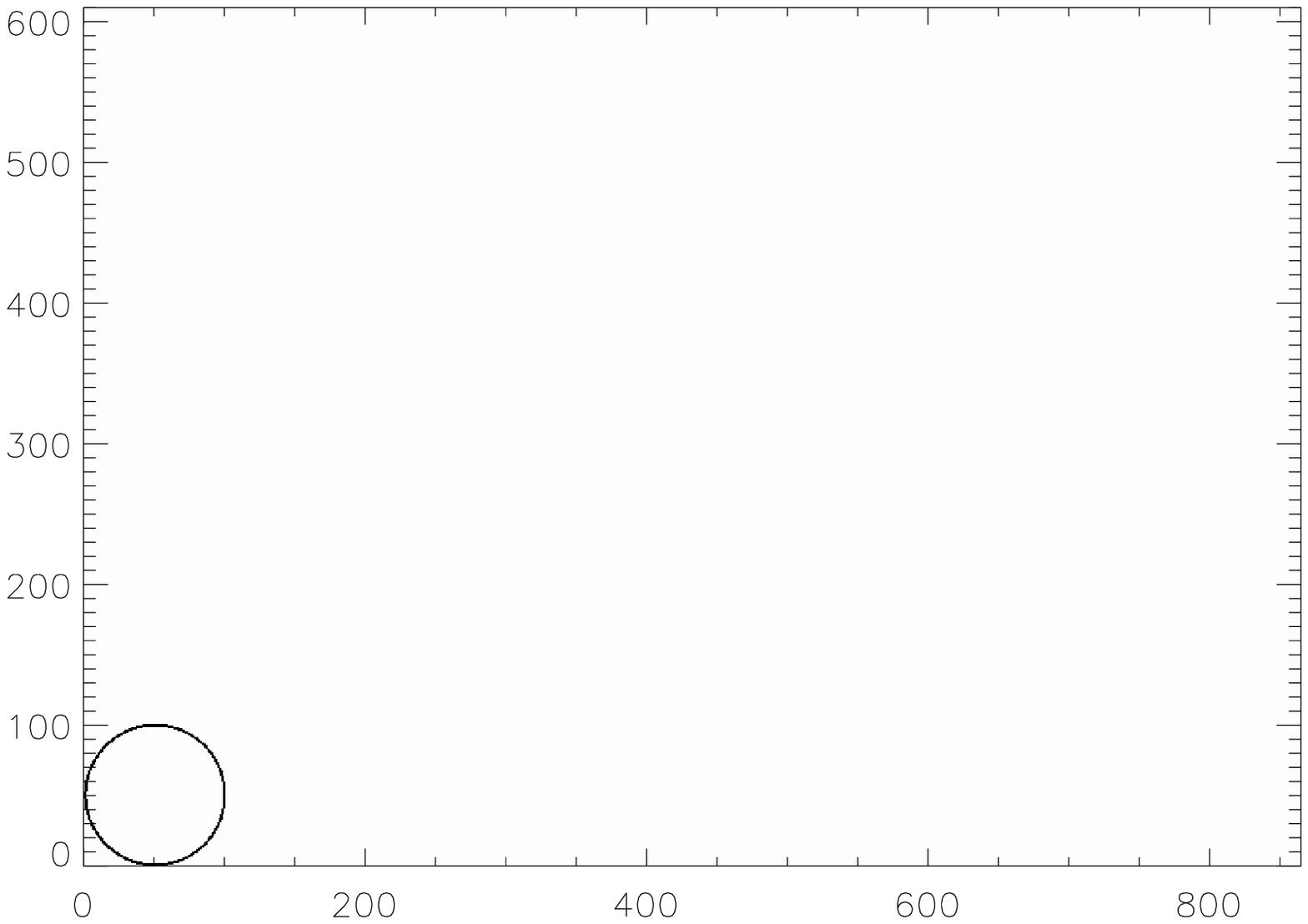}
}%
\\
\subfigure[]{ \centering
\includegraphics[width=84mm,angle=90]{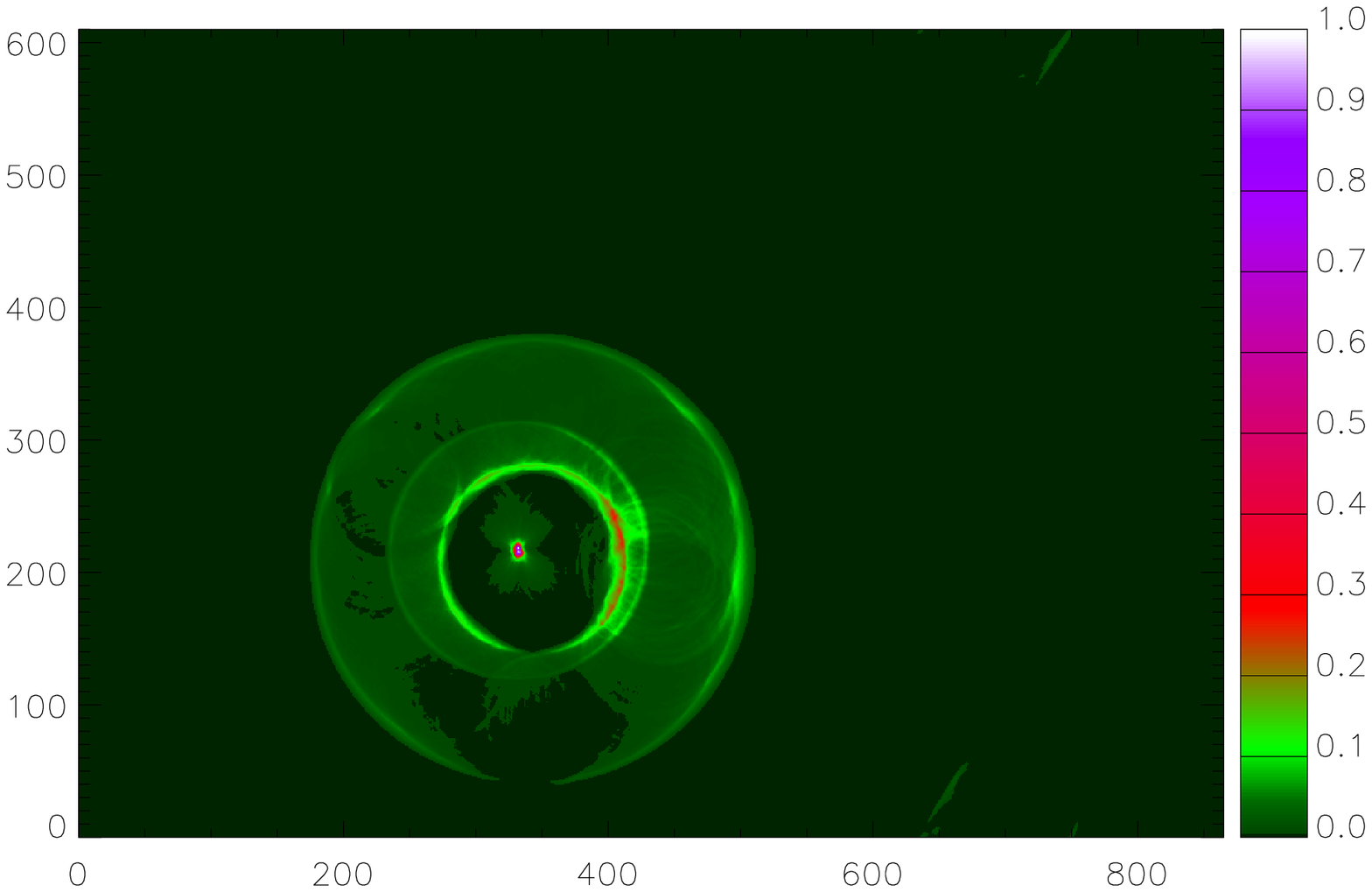}
}%
\subfigure[]{ \centering
\includegraphics[width=84mm,angle=90]{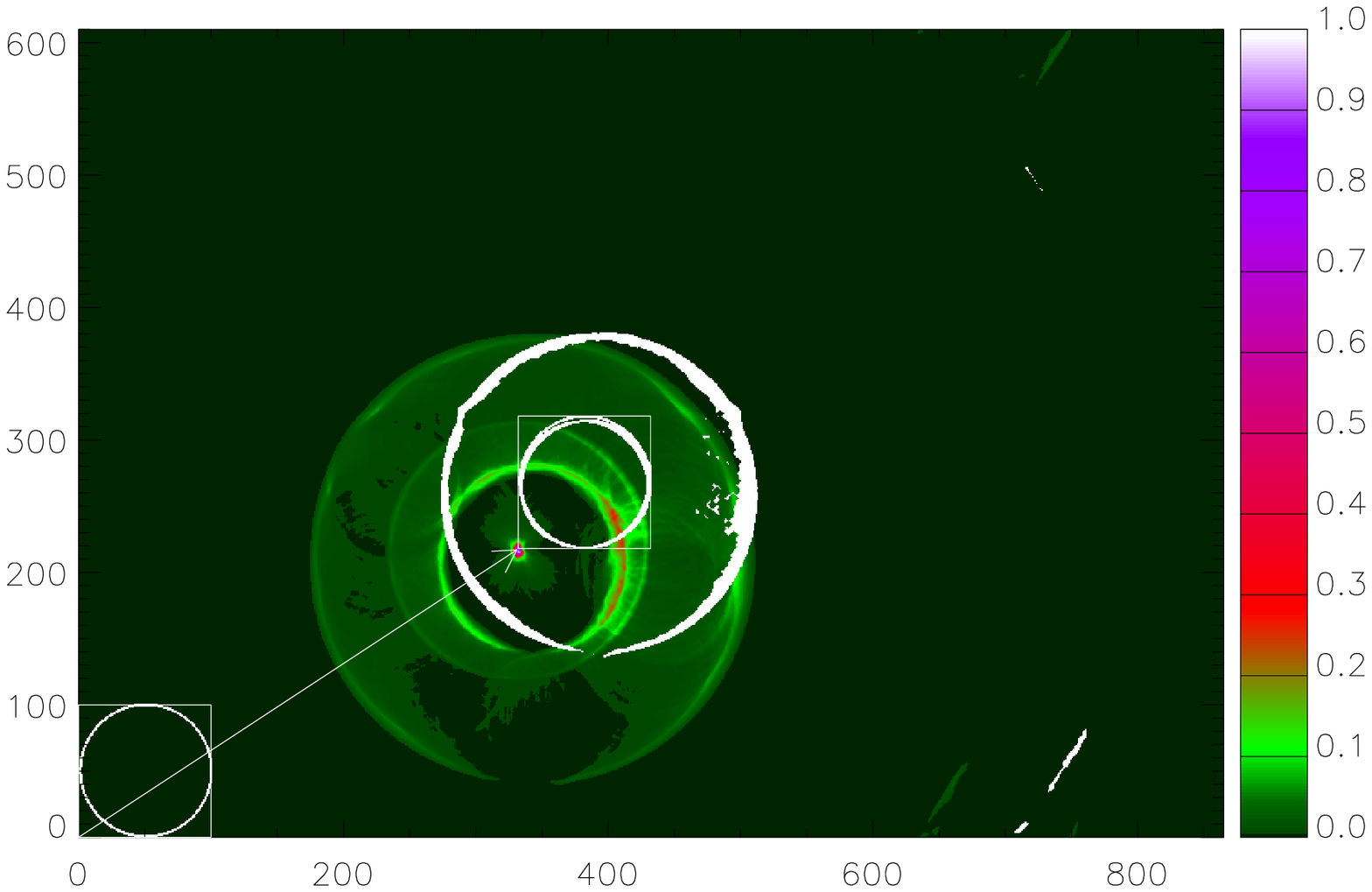}
}%
\caption{(a) is the padded edge image. (b) is the padded template.
(c) is the cross-correlation between (a) and (b). (d) is the
superimposed image of (a), (b) and(c).} \label{fig:7}
\end{figure*}

The subsequent step consists of extracting the edge of the solar
disk and of the ND4 filter from the respectively detected edges.
This step is achieved through image-matching based on
cross-correlation. For the solar disk, the angular diameter can be
obtained from the recording time of the image. Combining with the
focal length of the SBM, we can gain the theoretical size of the
solar diameter on the image. By taking into account the size of CCD
element, we finally get the solar diameter in pixels. Then the
template for cross-correlation can be constructed. Indeed, a
5-pixels width ring is constructed and adopted as the template
instead of a simplified circle template. The results are shown in
Fig.~\ref{fig:7}. The image and template are padded to avoid
wraparound error. The maximum of the resulted correlation-image
highlights the displacement with which the template should be
shifted. According to this recovered displacement and the size of
the template, the edge of the solar disk can be extracted from the
image reported in Fig.~\ref{fig:6}. Afterwards, by applying the
least-square approach we fit the circle using the edge points of the
solar disk to accurately estimate the coordinates of solar center
and radius. Similar procedures are performed on the edge of the ND4
filter.

{\color{blue} For the solar disk analyses, the cross-correlation technique might
give wrong displacement. In some situations, there is so much scattered light between the solar disk and the inner edge of the ND4 filter (diffraction ring) that the thresholding image from \textit{Otsu}'s Method might give ambiguous edges therein. Consequently, when the template is shifted to the position where the scattered light is evident, the correlation coefficient is significantly amplified. To overcome this, we proceed by calculating the convolution of the image-of-correlation and the corresponding
{\it Laplacian Kernel} function. It turns out that the correlation maximum of the defective image-of-correlation is effectively reduced after convolution. It is noted that in the case of the ND4 filters, there is no similar problem found.}

\subsection{Locating the position of the supporting arms}
\begin{figure*}
\subfigure[]{ \centering
\includegraphics[width=84mm,angle=90]{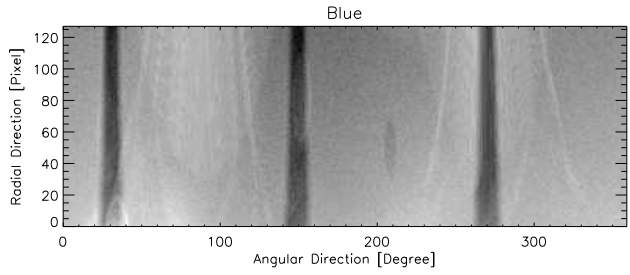}
}%
\subfigure[]{ \centering
\includegraphics[width=84mm,angle=90]{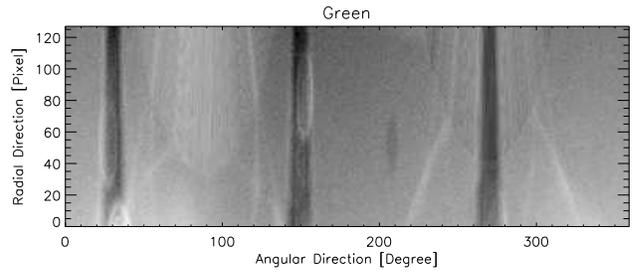}
}%
\\
\subfigure[]{ \centering
\includegraphics[width=84mm,angle=90]{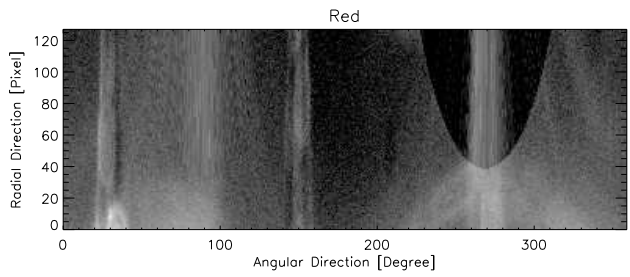}
}%
\subfigure[]{ \centering
\includegraphics[width=84mm,angle=90]{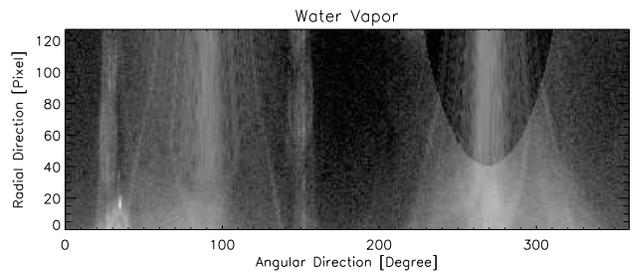}
}%
\caption{The expanding sky regions at different wavelengths. The
parabolic regions are the annular regions out of the image
constructed by interpolating with the points on the edge.}
\label{fig:10}
\end{figure*}

Considering the regions of the supporting arms' projections could
not be used for the sky brightness calculations, one needs hence to
get rid of these regions. For this purpose we test the
cross-correlation technique to locate the position of the supporting
arms. However, it can not be applied directly because the template
with invariable characteristics is hardly to be constructed.
Moreover, as long as the supporting arms are fixed on the ND4
filter, we can use the center of the ND4 filter as the pole to
expand the image in polar coordinate system. As shown in
Fig.~\ref{fig:10}, an annular region out of the diffraction ring is
isolated and expanded. The expanded image reveals a clear
distinctive feature, namely, three parallel strips with
$120^{\circ}$ away from each other.

\begin{figure*}
\subfigure[]{ \centering
\includegraphics[width=84mm,angle=90]{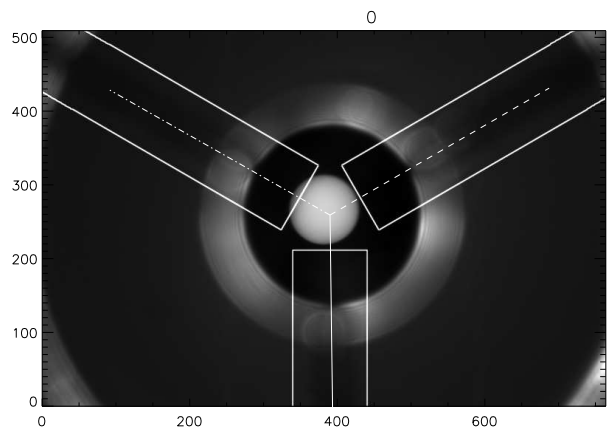}
}%
\subfigure[]{ \centering
\includegraphics[width=84mm,angle=90]{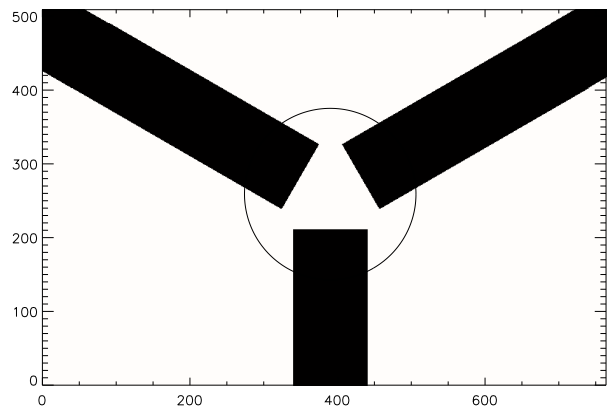}
}%
\\
\subfigure[]{ \centering
\includegraphics[width=84mm,angle=90]{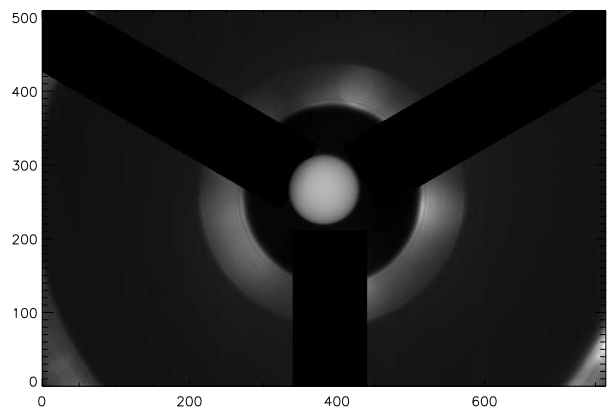}
}%
\subfigure[]{ \centering
\includegraphics[width=84mm,angle=90]{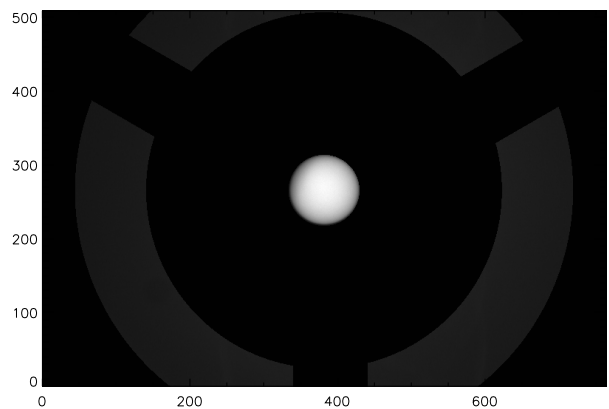}
}%
\caption{(a) is the SBM image. (b) is the mask. (c) is the product
of (a) and (b). (d) shows the solar disk and the sky regions.}
\label{fig:11}
\end{figure*}

Worthy of note, for convenience we inverse the values of the
expanded image. Because the projections of the supporting arms in the expanded images are relatively fixed and parallel, a two-column template can work well enough. Moreover, the sky brightness is not uniform at the angular direction, the template constructed adopting two {\it Gaussian} functions is more suitable than using three {\it Gaussian} functions. Due to our concerning only the positions of the supporting arms on the angular direction, the template needs not to be designed too high. On consideration of the efficiency, we eventuality adopt a template consists of two {\it Gaussian} functions aligned with their maximums at a distance of $120^{\circ}$ and with a width of 32 pixels. The results are illustrated in Fig.~\ref{fig:11}.

\begin{figure*}
\centering
\includegraphics[width=168mm,angle=90]{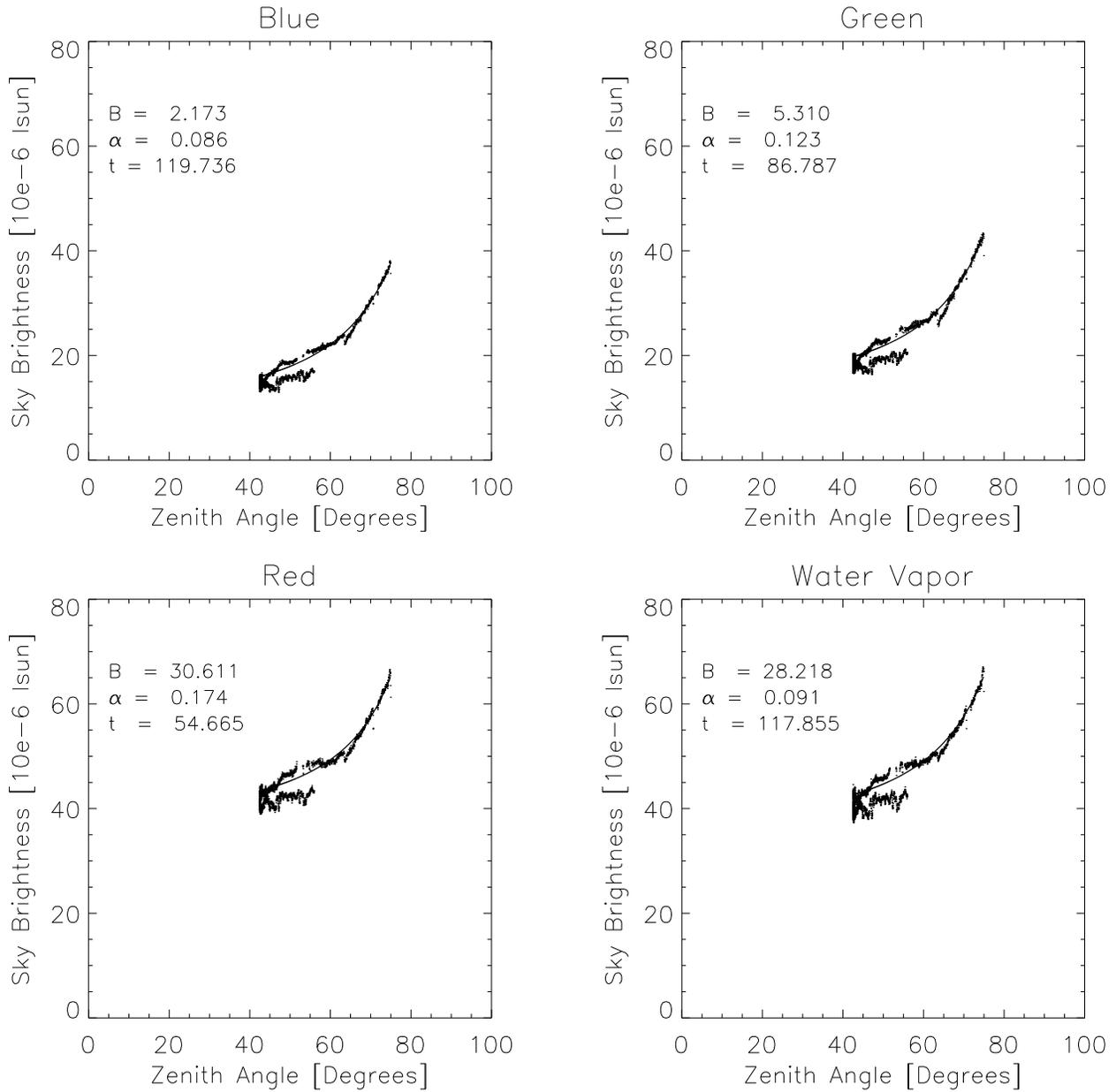}
\caption{Fitted sky brightness to the formula
$S(Z)=B-\alpha R\cos Z +\alpha\sqrt{R^2\cos^2Z+2Rt+t^2}$ in four wavelengths. Blue,
green, red and water vapor correspond to 450 nm, 530 nm, 890 nm and
940 nm, respectively.} \label{fig:12}
\end{figure*}

The SBM instrument measures the solar disk and sky brightness at
four wavelengths, including 450 nm, 530 nm, 890 nm and one at the
water vapor absorption band of 940 nm. The data acquisition system
is collecting images at four wave bands in the sequence of
wavelengths. The scattered light of the sky might be found weaker
than that of the supporting arms especially in some cases when the
sky brightness is too weak. An example is depicted in
Fig.~\ref{fig:12}. In this situation, the images at 890 and 940 nm
will present different characteristics from those at 450 and 530 nm.
Thus, the above template is not always suitable for all the four
wave bands. The SBM instrument requires approximately a few tens of
seconds to acquire a set of data which contains four images at
different wavelengths. The positions of the supporting arms are not
significantly changing in a given data set. For each data set, our
procedure consists of computing the positions from 450 nm, adopting
cross-correlation method, henceforth apply them on the other
different three wavelengths.
\subsection{Calculation of the sky brightness}
\begin{figure*}
\subfigure[]{
\includegraphics[width=84mm,angle=90]{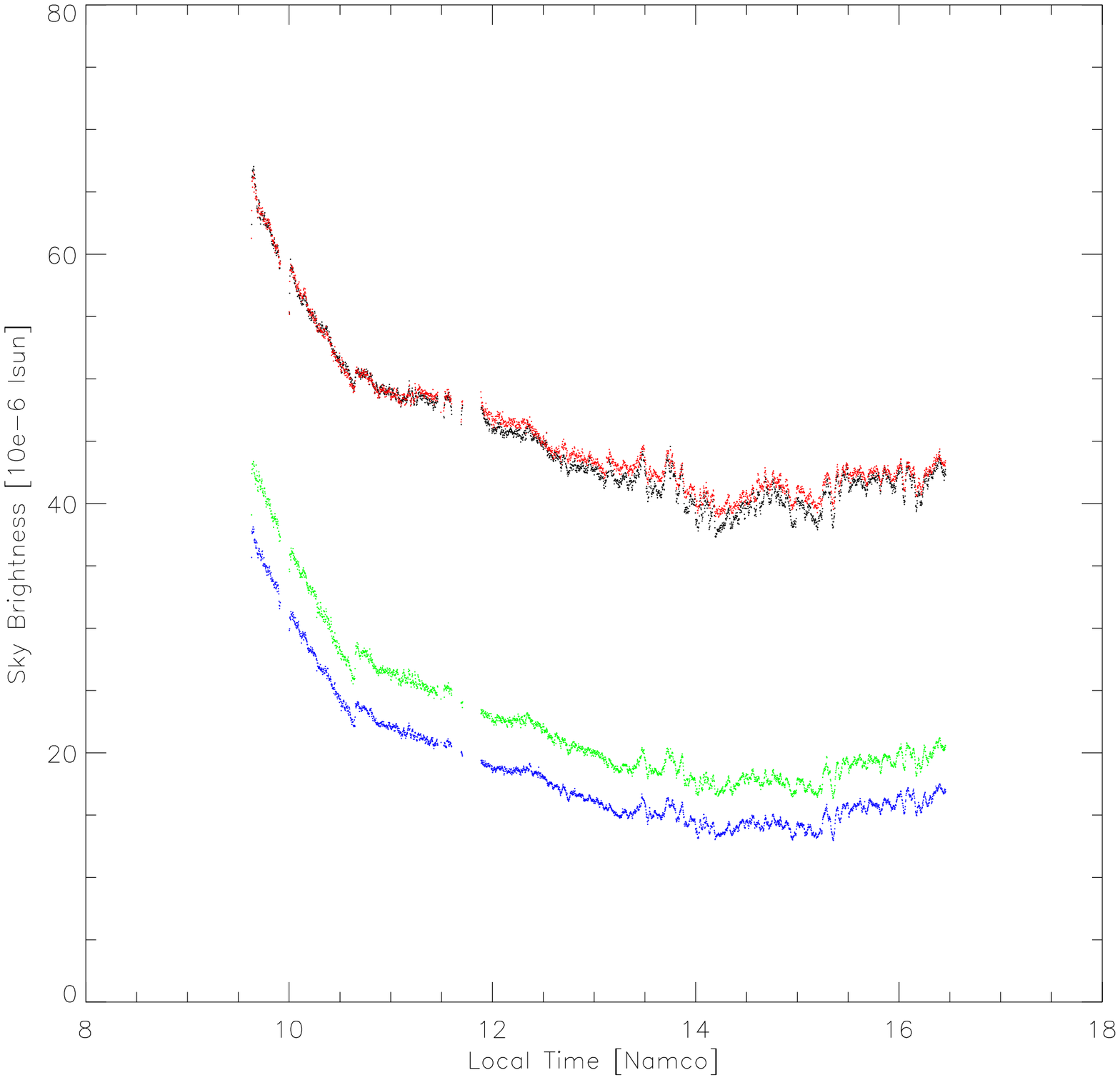}
}%
\subfigure[]{
\includegraphics[width=84mm,angle=90]{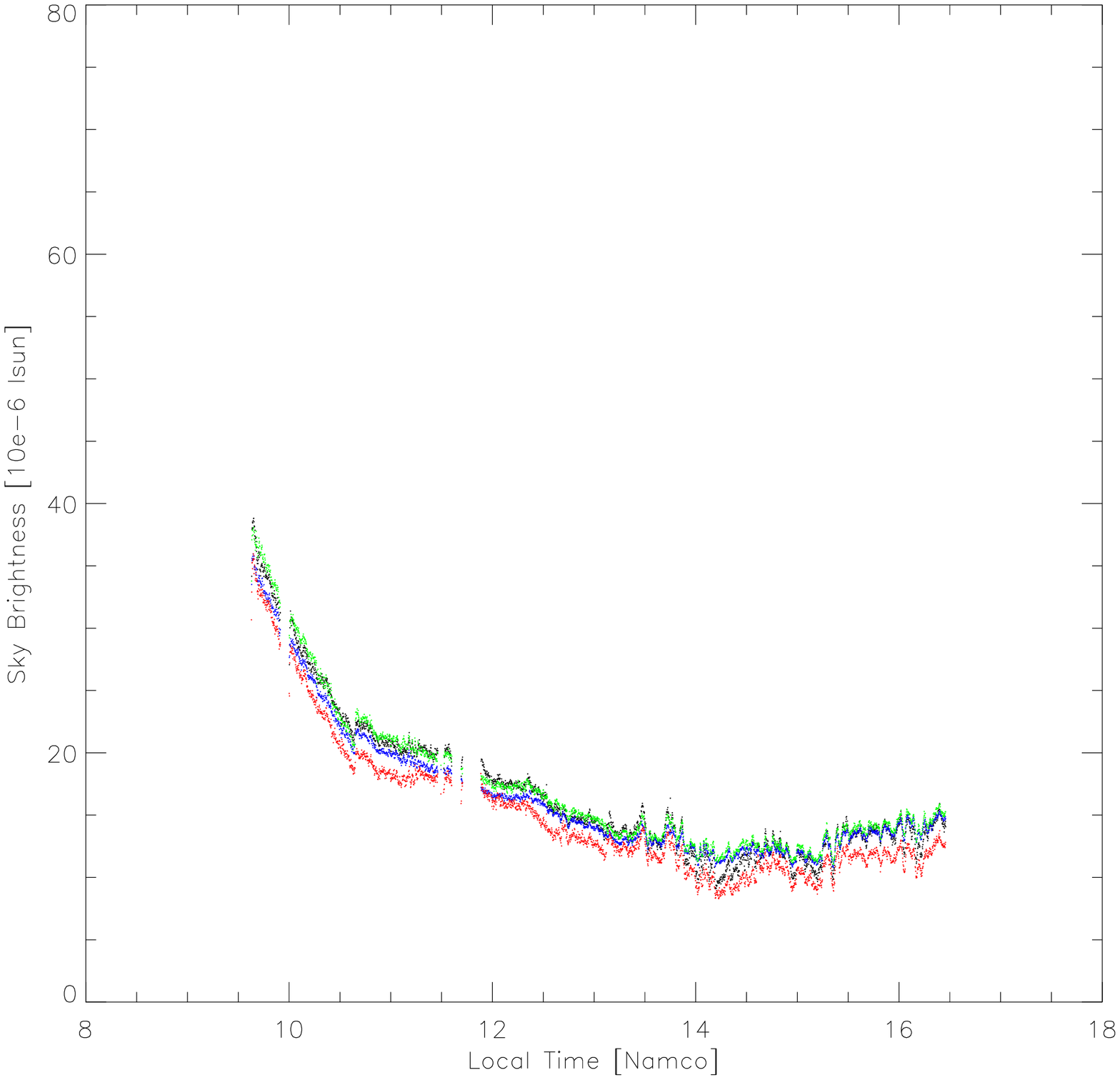}
}%
\caption{The variations of the sky brightness with the local time at
Lake Namco before and after the calibration of the scattered light.}
\label{fig:13}
\end{figure*}

The sky brightness is usually measured as follows,
\begin{equation}
S=\frac{I_{sky}}{I_{sun}}, \label{eq:5}
\end{equation}
where $I_{sky}$ is the average intensity at the region several solar
radii away from the solar disk and $I_{sun}$ is the true intensity
at the solar disk center before weakened by the ND4 filter. We note
that the scattered light of the SBM is not included in the above
formula and the scattered light will be discussed in the next
section. In our measurement, the region corresponding to $I_{sky}$
is between 5.0 to 7.0 solar radii excluding the projections of the
supporting arms. For this purpose, we construct a mask based on the
angular coordinates of the positions and the central coordinates of
the ND4 filter. Then the mask is multiplied by the image of SBM.
Next, we get rid of the region out of 7.0 solar radii and in 5.0
solar radii from the masked image according to the fitting position
of the solar center and solar radius. The results are reported in
Fig.~\ref{fig:13}.

The solar intensity $I_{sun}$ is usually computed in two ways: a
central solar intensity is computed within $0.1 R_{sun}$, or a mean
solar intensity is computed using all pixels out to $1 R_{sun}$
\citep{Penn2004}. Generally, the first way is adopted in former
works \citep{Penn2004,Lin2004,Liu2012b}. However, the solar center
in the image does not necessary coincide with the physical center of
the Sun. This is mainly because the tube of SBM is not always
perfectly parallel to the light ray during observations. The
observers must adjust the position of the SBM after tracking the Sun
for a period of time, which is about thirty minutes to two hours
based on the desired alignment accuracy. Here we stress once again
that the portability of the device is of great importance and we
should avoid spending too much time with the alignment issue in a
site survey. As a consequence, the mean solar intensity near the
disk center may seem to "sharply" change with time. In order to
avoid this possible inconsistency, we adopt the intensity at the
centroid of the solar disk in the image for the sky brightness
calculations.

\begin{table*}
\centering
\begin{tabular}{r|l|c|c|c}
\hline
No. & Pass band & Sky brightness & Recording time (UT) & Zenith angle($^{\circ}$)\\
\hline
0 & Blue  & 35.7 & 2013-10-23T01:37:53 & 74.879\\
1 & Green & 39.1 & 2013-10-23T01:37:50 & 74.889\\
2 & water & 62.4 & 2013-10-23T01:37:55 & 74.872\\
3 & Red   & 61.3 & 2013-10-23T01:37:45 & 74.905\\
4 & Blue  & 37.8 & 2013-10-23T01:38:02 & 74.850\\
5 & Green & 42.4 & 2013-10-23T01:37:59 & 74.859\\
6 & water & 66.2 & 2013-10-23T01:38:04 & 74.843\\
7 & Red   & 63.5 & 2013-10-23T01:37:57 & 74.866\\
...&   ...&     ...&              ...&    ...\\
...&   ...&     ...&              ...&    ...\\
5121 & Blue  & 13.6 & 2013-10-23T07:08:30 & 45.859\\
5122 & Green & 16.9 & 2013-10-23T07:08:27 & 45.855\\
5123 & water & 38.8 & 2013-10-23T07:08:32 & 45.862\\
5124 & Red   & 40.0 & 2013-10-23T07:08:25 & 45.852\\
5125 & Blue  & 13.4 & 2013-10-23T07:08:39 & 45.873\\
5126 & Green & 17.0 & 2013-10-23T07:08:36 & 45.868\\
5127 & water & 38.7 & 2013-10-23T07:08:41 & 45.876\\
5128 & Red   & 40.0 & 2013-10-23T07:08:34 & 45.865\\
...&   ...&     ...&                  ...&    ...\\
...&   ...&     ...&                  ...&    ...\\
10414 & Blue  & 18.0 & 2013-10-23T04:30:26 & 47.058\\
10415 & Green & 22.1 & 2013-10-23T04:30:23 & 47.063\\
10416 & water & 44.4 & 2013-10-23T04:30:28 & 47.054\\
10417 & Red   & 45.2 & 2013-10-23T04:30:21 & 47.066\\
10418 & Blue  & 18.1 & 2013-10-23T04:30:35 & 47.042\\
10419 & Green & 21.8 & 2013-10-23T04:30:32 & 47.047\\
10420 & water & 45.0 & 2013-10-23T04:30:37 & 47.038\\
10421 & Red   & 44.8 & 2013-10-23T04:30:30 & 47.051\\
\hline
\end{tabular}
\caption{Sample results based on our automatic procedure on the raw
data set taken on October 23, 2013 at Namco Lake in Tibet
($31^{\circ}$N, $90^{\circ}$E). The unit for sky brightness is in
millionth of the corresponding solar disk center intensity.}
\label{tab:1}
\end{table*}

\section{Results and Analysis}
In this part we highlight our analysis and approach; based on the
techniques described in the previous sections; applied to the data
set acquired on October 23, 2013 at {\it Namco Lake} in Tibet. Our
procedure takes approximately four hours to automatically process
the 10422 data with a normal PC with a dual-core processor (Intel
Core i3-2350M CPU at 2.30GHz). The results are exported in a
table-format output containing: the wave bands, recording time, sky
brightness and solar zenith angle, respectively. These output
results can subsequently be used for the evaluation of a series of
important physical parameters. A short list sample of the results is
reported in Table~\ref{tab:1}. Note that those raw data are not
listed sequentially in time. Sorting the results by Julian date
counts, due to the recording time, is an effective and necessary
step for further calculations.

During the coronagraph site survey, the performance of the SBM
instruments in long-time field work may cause unexpected parameter
changes, such as increased scattered light levels for different wave
bands. It is hence necessary to calibrate the scattered light levels
frequently. According to \citet{Mar1990}, the normalized sky
brightness including a constant instrumental scattered light can be
expressed as follows:
\begin{equation}
\frac{I_{sky}}{I_{sun}}=\phi_{\kappa}M+B, \label{eq:6}
\end{equation}
where $\phi_{\kappa}$ is the atmospheric scattering function, $B$ is
an instrumental constant which depends on wavelength and $M$ is the
air mass. Following \citet{Lin2004}, the air mass column can be calculated
as a function of zenith angle under the uniform curved atmospheric
model:
\begin{equation}
M_c=-R\cos Z +\sqrt{R^2\cos^2Z+2Rt+t^2}, \label{eq:7}
\end{equation}
where $Z$ presents the zenith angle, $R$ equals the sum of the altitude and the radius of the Earth, and $t$ is the thickness of the atmosphere. Then we can get:
\begin{equation}
S(Z)=B-\alpha R\cos Z +\alpha\sqrt{R^2\cos^2Z+2Rt+t^2}
\label{eq:8}
\end{equation}
Therefore, we can fit the measured sky brightness $S(Z)$ as a
function of $Z$ to the above formula, in which $B$, $\alpha$
and $t$ are three free parameters in the fitting algorithm.

Fig.~\ref{fig:14} displays the variations of the sky brightness
measured with time in all four wavelengths throughout the course of
a day. The fact that the measured sky brightness in the morning and
in the afternoon, with the same zenith angles, can not overlap with
each other, implies that the atmospheric conditions are not stable
throughout the day. \citet{Lin2004} did not suggest using this kind
of data to fit the model. However, to obtain the fitted sky
brightness we decide on adopting only the data during the morning
time, {\it i.e.}, the first half of the observations with better sky
condition. The calibration parameters of the scattered light are
2.06, 5.13, 30.26 and 28.06 millionths for the blue, green, red and
water vapor band-passes, respectively, obviously higher than those
values reported in \citet{Liu2012b}. The increased scattered-light
levels indicate that the black coating material inside the SBM tube
should have been seriously got faded after long-term sunshine
exposed to high-energy radiation in high altitude environment. This
problem will be resolved soon by coating the SBM tube regularly in
the factory. The coefficients of the ND2 filters show little changes
since we have tested them yearly. The variations of the sky
brightness with the local time at {\it Namco Lake} before and after
the calibration of the scattered light are shown in
Fig.~\ref{fig:15}.

Another possible application of our procedure is for guiding and
tracking the Sun with high precision because our methodology can
accurately derive the Sun's center coordinates. Usually, a telescope
can track an object using an extra guiding system composed by a tube
parallel to the main telescope and an individual CCD imager. For
field observation such as our site survey, the portable SBM
instruments were not designed to be equipped with extra guiding
system parallel to the main telescope tube. For most guiding systems
for solar telescopes with full-disk observing mode, the Sun's
position for tracking is usually derived from the centroid of solar
image. In one test with a white-light solar telescope of 100 mm
aperture, we found the centroid of the Sun could deviate from that
position, based on least-square fitting calculations, for about 50
pixels. In that test, the commercial CCD size is 5202 by 3454 pixels
and the data are white-light full-disk solar images with solar
diameter of about 3050 pixels. Moreover, our algorithm is more
robust compared with the centroid method, because the fitting
doesn't require all the points of the Sun. When the Sun is partly
covered by clouds, our algorithm can give the right results while
the centroid method would not. Interestingly, all the solar
telescopes with the full-disk mode can hence benefit from our
algorithm and they can track the Sun without an extra guiding
system.

\section{Conclusions}
For the solar site survey data analysis, we have developed one
automatic procedure for the SBM instruments. A detailed description
of the procedure is presented. The method can calculate the sky
brightness, including identifying the solar disk and the sky areas
around it as well as the excluded part of the supporting arms. These
processes are implemented on our data acquired by different SBM
instruments, including those made in China or lent from USA, during
the site survey project in Western China. These results are
significant and highly encouraging for further future efforts and
investigations, particularly in deriving a variety of important
physical parameters such as extinctions, aerosol content and
precipitable water vapor content. Furthermore, the Sun's accurate
center coordinates derived in our procedure can be applied for
tracking the Sun without an extra guiding system. Indeed, we are in
a consulting phase with our engineers for a possible near-future
practical realization of the project.

The preliminary analysis applying the results has been performed and
discussed for the assessment of the levels of the scattered light.
When we compare the scattered light levels obtained from this work
with those taken 2 years ago \citep{Liu2012b}, we notice significant
increase in every wavelength. This should be due to the functional
degradation of the dark material coating inside the SBM tube which
has been used after long-term sunshine exposure in the environment
of the Tibet plateau. The new results about the scattered light
levels will be reported after the tube is coated again in the near
future. However, the robotization of our method for SBM data
analysis has been tested and confirmed with high speed based on
large data sample taken in the year of 2013 for various candidate
sites, without any artificial intervention such as changing some
parameters according to different SBM instruments or weather
conditions as before.

\section*{Acknowledgments}
We thank Dr. Matt Penn for his helpful comments to improve the manuscript. This work was supported
by the Natural Science Foundation of China (NSFC) under grants 10933003, 11078004, 11073050,
11203072, 11103041, 11178005 and U1331113, the National
Key Research Science Foundation (2011CB811400), and the Open
Research Program of Key Laboratory of Solar Activity of National
Astronomical Observatories of Chinese Academy of Sciences
(KLSA201204). AE is partly supported by Chinese Academy of Sciences
Fellowships for young international scientists, grant 2012Y1JA0002.
YL is partly supported by the Visiting Professor Program of King
Saud University.



\begin{figure*}
\subfigure[]{
\includegraphics[width=84mm,angle=90]{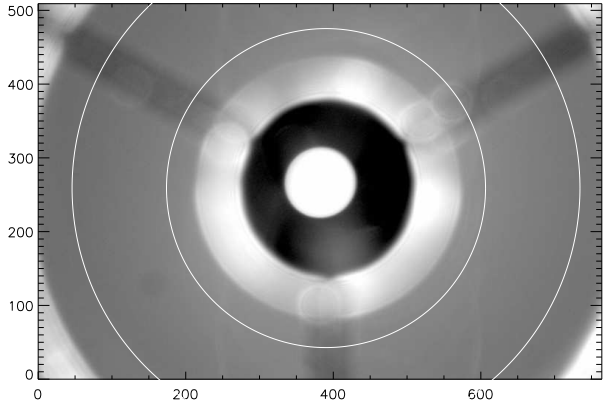}
}%
\subfigure[]{
\includegraphics[width=84mm,angle=90]{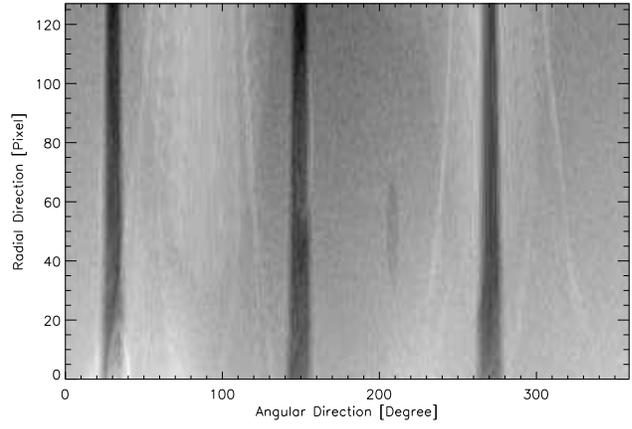}
}%
\caption{Taking the annular region from (a) and expanding it in
polar coordinate.} \label{fig:8}
\end{figure*}

\begin{figure*}
\centering \subfigure[]{ \centering
\includegraphics[width=84mm,angle=90]{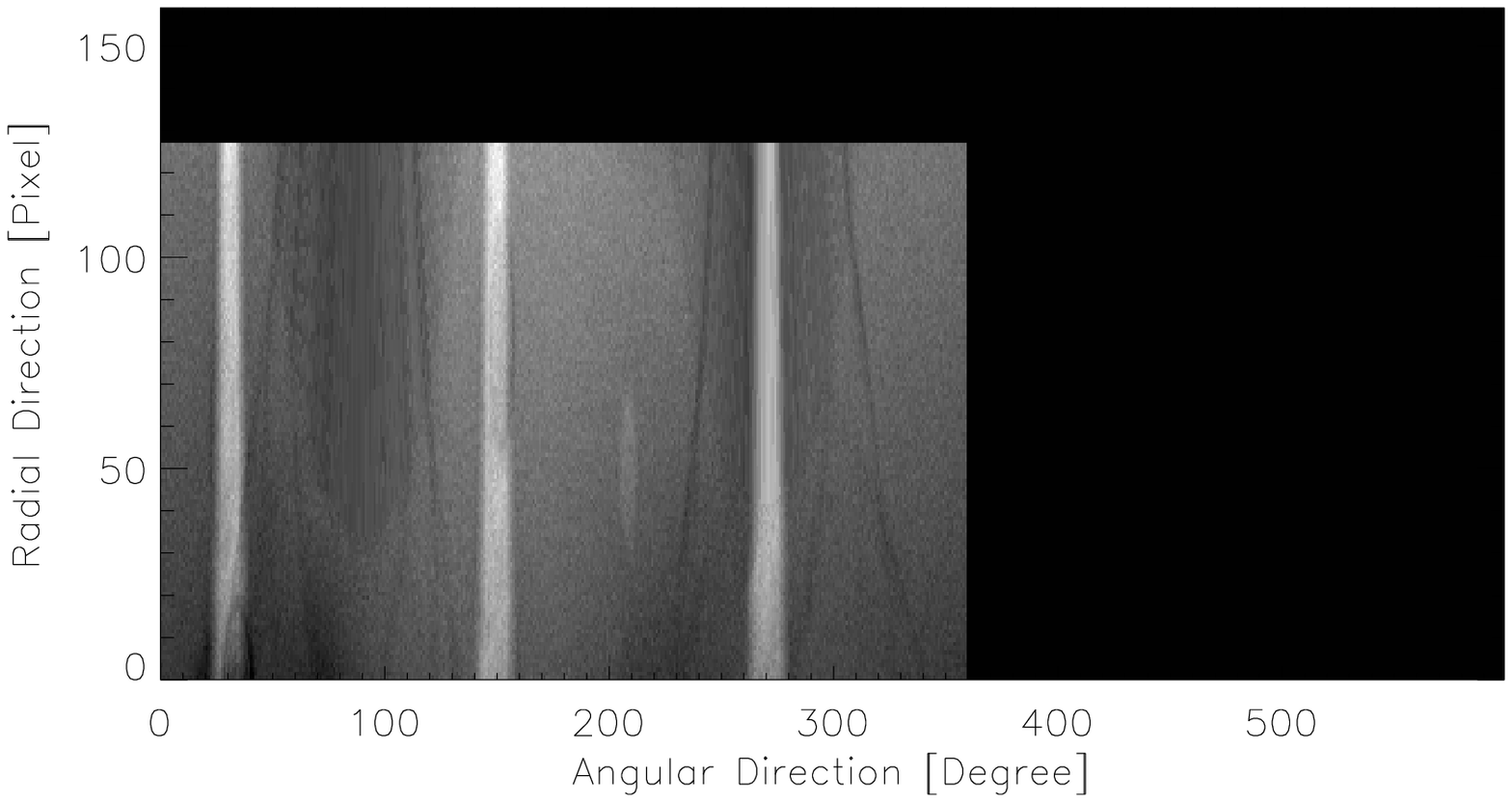}
}%
\subfigure[]{ \centering
\includegraphics[width=84mm,angle=90]{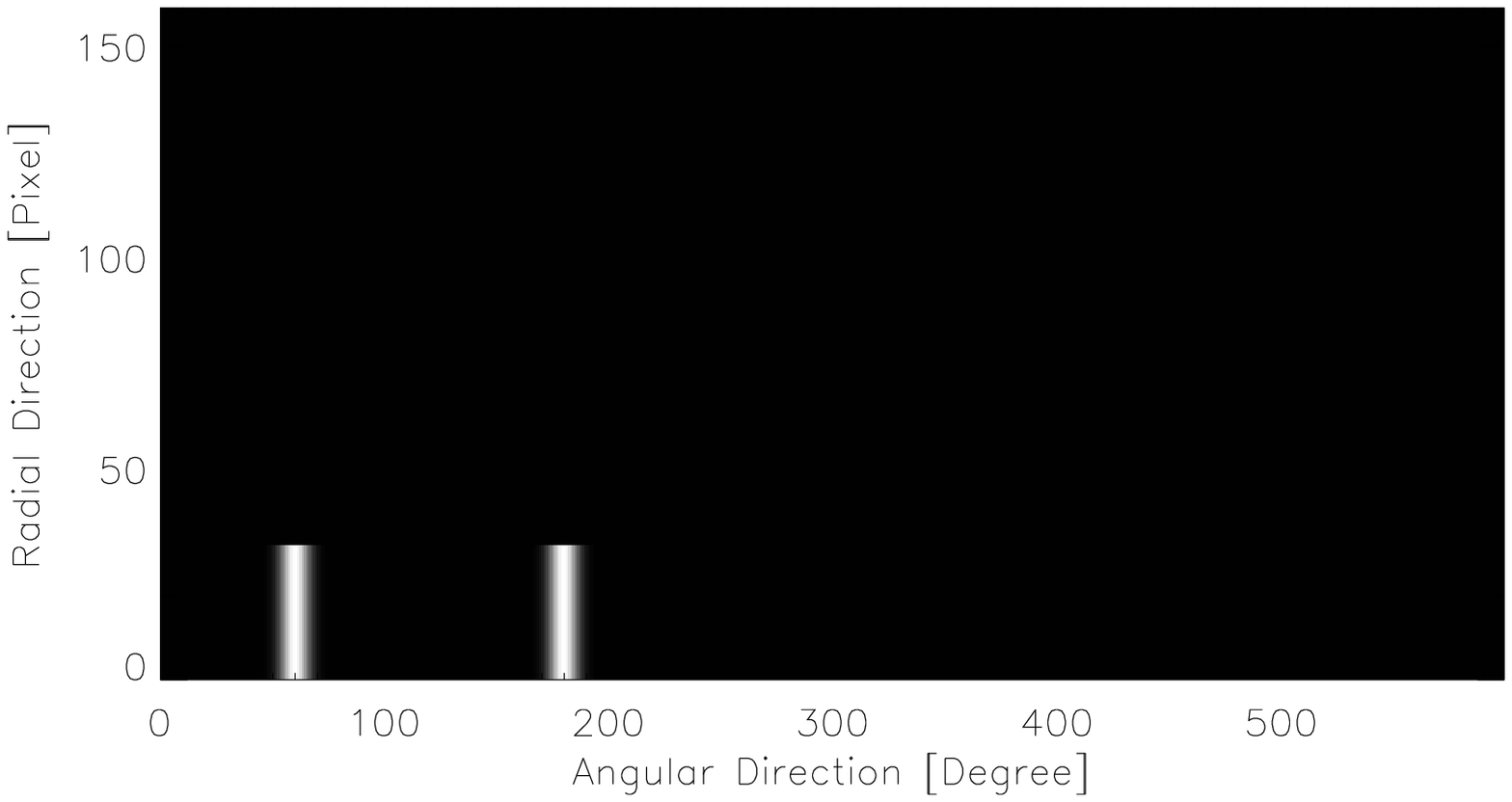}
}%
\\
\subfigure[]{ \centering
\includegraphics[width=84mm,angle=90]{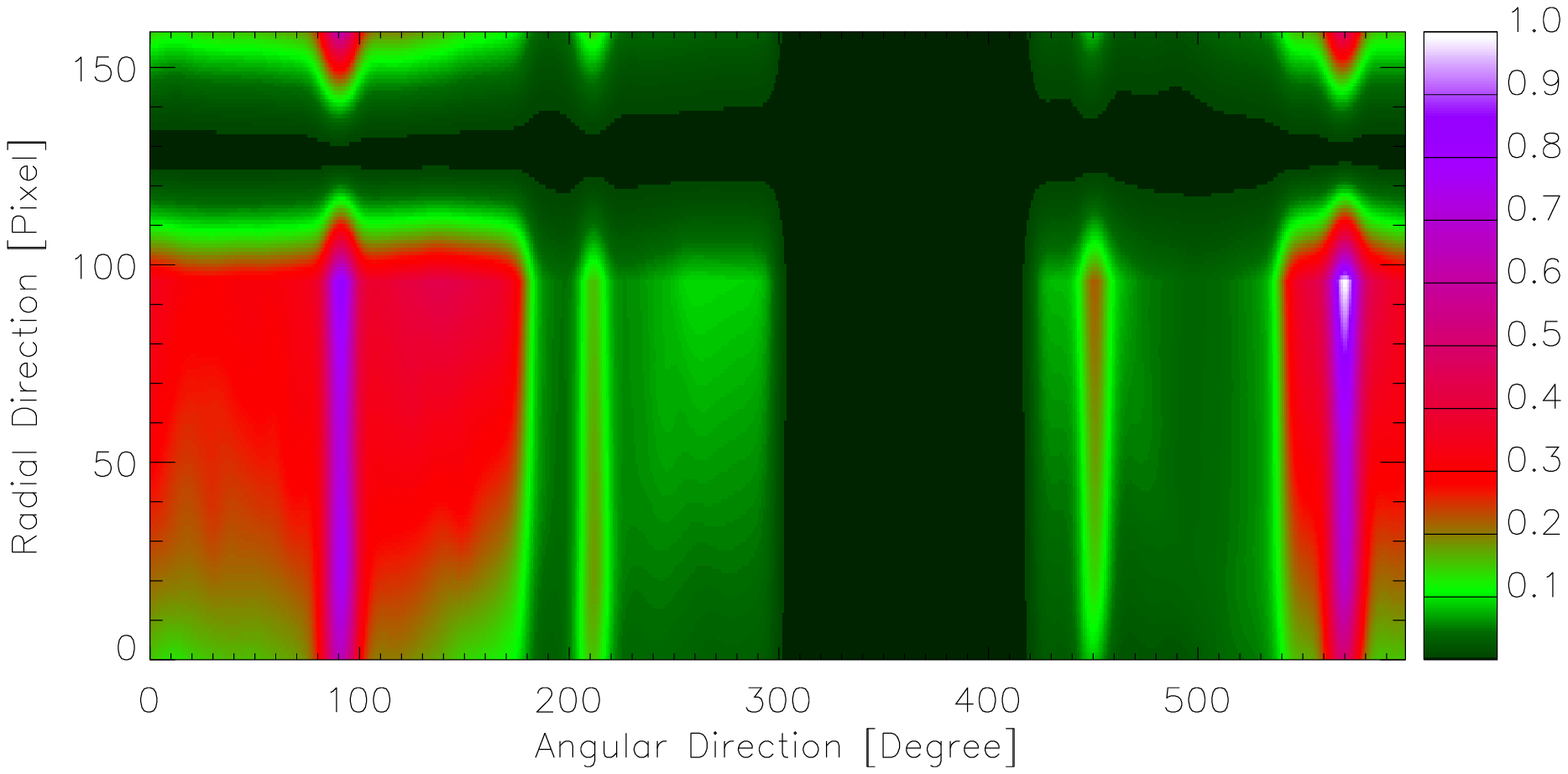}
}%
\subfigure[]{ \centering
\includegraphics[width=84mm,angle=90]{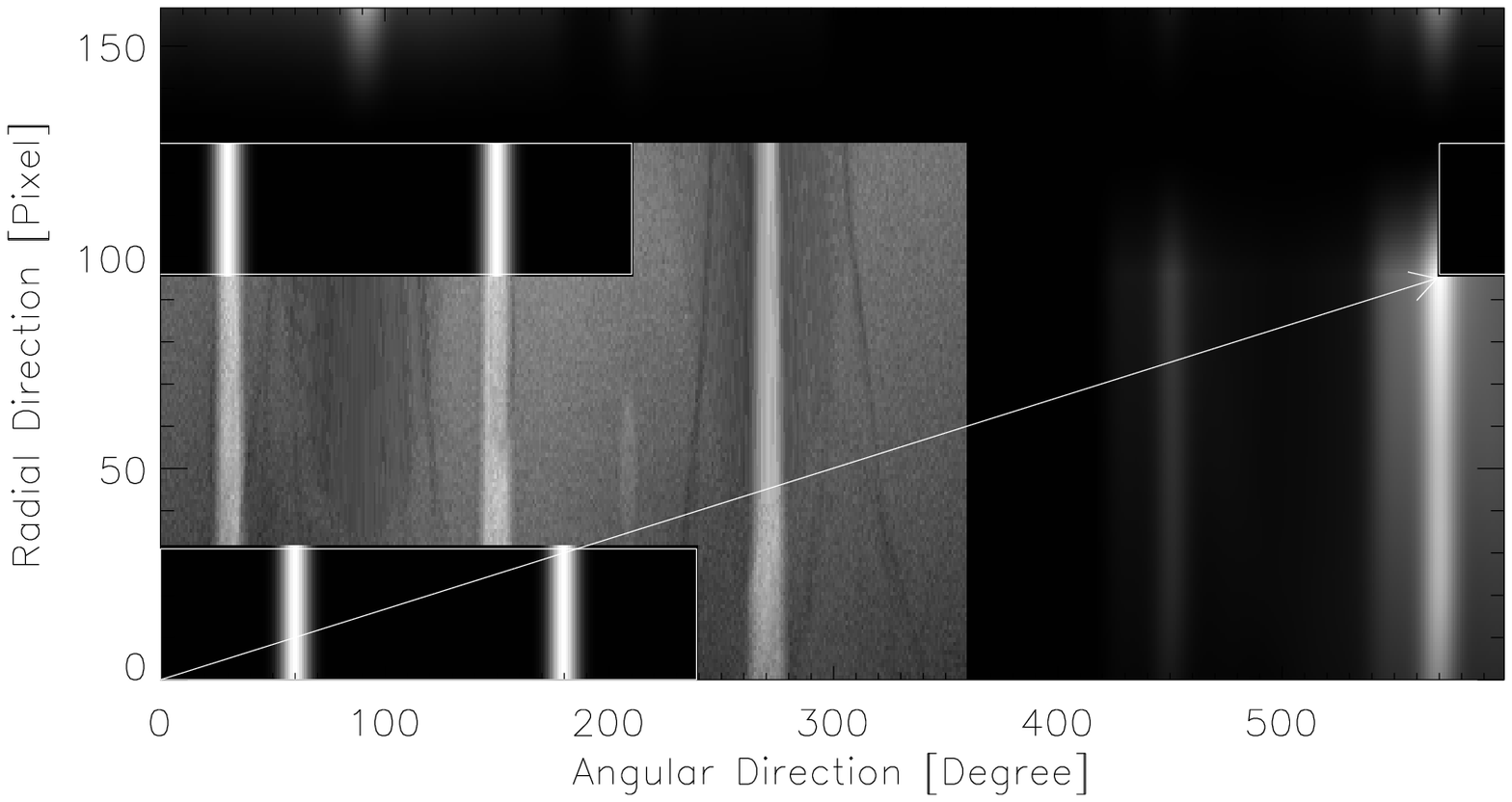}
}%
\caption{(a) is the padded edge image. (b) is the padded template.
(c) is the cross-correlation between (a) and (b). (d) is the
superimposed image of (a), (b) and(c).} \label{fig:9}
\end{figure*}



\end{document}